\documentclass[aps,pre,twocolumn,groupedaddress,showkeys,showpacs]{revtex4}
\usepackage{graphicx}
\usepackage{amssymb}
\usepackage{amsmath}

\def\tri {\bigtriangleup}

\begin{document}

\title{Betti number signatures of homogeneous Poisson point processes.} 


\author{Vanessa Robins}
\email{Vanessa.Robins@anu.edu.au}
\affiliation{Department of Applied Mathematics, Research School of Physical Sciences, The Australian National University, Canberra ACT 0200, Australia}

\date{\today}

\begin{abstract}
The Betti numbers are fundamental topological quantities that describe the $k$-dimensional connectivity of an object: $\beta_0$ is the number of connected components and $\beta_k$ effectively counts the number of $k$-dimensional holes. 
Although they are appealing natural descriptors of shape, the higher-order Betti numbers are more difficult to compute than other measures and so have not previously been studied \emph{per se} in the context of stochastic geometry or statistical physics. 

As a mathematically tractable model, we consider the expected Betti numbers per unit volume of Poisson-centred spheres with radius $\alpha$.  
We present results from simulations and derive analytic expressions for the low intensity, small radius limits of Betti numbers in one, two, and three dimensions.  The algorithms and analysis depend on alpha-shapes, a construction from computational geometry that deserves to be more widely known in the physics community. 
\end{abstract}

\pacs{02.50.Ey Stochastic processes; 02.40.Re Algebraic topology; 05.10.Ln Monte Carlo studies}
\keywords{Topological invariants, Betti numbers, Euler characteristic, Poisson-Boolean model, alpha-shapes, continuum percolation.}

\maketitle


%%%%%%%%%%%%%%%%%%%%%%%%

\section{Introduction}

Topological measures of shape are finding increasing use in the study of point or coverage processes and the characterisation of complex three-dimensional structures
\cite{MeckeBook2}. 
This is because topology is independent of geometry, and so both types of information are necessary to fully characterise spatial structure~\cite{RobinsChapter02,RobinsThesis}.  
The most commonly studied topological invariants are the number of connected components (the zeroth order Betti number, $\beta_0$) and the Euler characteristic ($\chi$, the zero-dimensional Minkowski measure from integral geometry).  
This paper also investigates $\beta_1$ and $\beta_2$, the higher-order Betti numbers that count the number of independent handles (non-contractible loops) and enclosed voids. 

The Betti numbers are closely related to the Euler characteristic via the Euler-Poincar\'e formula: $\chi = \beta_0 - \beta_1 + \ldots - \beta_m$.  For subsets of $\mathbb{R}$, the Euler characteristic is exactly the number of components.  In $\mathbb{R}^2$, there are only two independent quantities from the three, since $\chi = \beta_0 - \beta_1$.  
Thus, as $\chi$ and $\beta_0$  are already well-known quantities in statistical physics, the higher-order Betti numbers give intrinsically new information only in dimensions three and higher.   
Nonetheless, it is instructive to study the Betti numbers directly in both two- and three-dimensions as they give a more direct description of the topology than the Euler characteristic. 
For example, recent work on the 2D Griffiths' model has used Betti numbers of the different states  to characterise the phase transition~\cite{Blanchard06}. 

In this paper, we present analysis and simulations of the Betti number signatures of Poisson point patterns. 
The Poisson point process is the most widely studied model in stochastic geometry and is frequently used as a null hypothesis for comparison with physical systems~\cite{StoyanBook}.  
In general, a \emph{signature} for a point pattern is defined by attaching spheres of radius $\alpha$ to each point and computing some quantity of interest as a function of $\alpha$. 
Thus, the Betti number signatures contain both topological \emph{and} geometric information about the distribution of points in space.  
In applications, such signature functions can be used to detect differences between simulations and physical data, or to provide insight into the physical processes that generated a particular distribution of points. 
Example applications will be the topic of a future paper. 

We give a brief overview of the simulation of Poisson point processes in Section~\ref{PoissonPointProcesses}.
Our computation and analysis of the Betti number signatures use \emph{alpha-shapes}~\cite{Edels83,Edels94} --- a construction from computational geometry that is dual to the union of spheres of radius $\alpha$.   The alpha-shape is a subcomplex of the Delaunay triangulation of a set of points, so we can draw on extensive results about Delaunay complexes of Poisson-distributed points. 
We summarise the alpha-shape and Betti number algorithms in Sections~\ref{BettiNumbers} and~\ref{PeriodicBoundary}.  
Results of the simulations are presented in Section~\ref{Results}.  The final section (\ref{Theory}) of the paper gives derivations of the low-intensity small-radius behaviour of the Betti numbers of Poisson-distributed spheres.

%%%%%%%%%%%%%%%%%%%%%%%%

\section{Simulation methods}

\subsection{Poisson point processes}
\label{PoissonPointProcesses}

A Poisson point process in $\mathbb{R}^d$ with constant intensity $\lambda$ is easily simulated in the unit $d$-cube by generating $N$ points with $d$ coordinates chosen from a uniform random distribution on $[0,1]$.  The number of points, $N$, is a random variable generated from a Poisson distribution with mean $\lambda$,
\[  
Pr(N = n) = \lambda^n e^{-\lambda}/n! 
\] 
For large values of $\lambda$, Poisson distributed numbers are well approximated by a normal (Gaussian) distribution with mean $\lambda$ and standard deviation $\sqrt \lambda$, i.e.,
\[
 Pr(N = n) =  \int_{n-0.5}^{n+0.5} f(x) dx ,  
 \]
 where $f(x)$ is the normal probabiltiy density function
 \[
 f(x) =  \frac{1}{\sqrt{2\pi \lambda}} \exp[ -(x-\lambda)^2 / 2\lambda ] . 
\]
Given a realisation of a Poisson point process in the unit $d$-cube, label the $N$ points $X_1, X_2, \ldots, X_N$ and place identical balls of radius $\alpha$ centred at each point, $B_d(X_i,\alpha)$. The Betti number signatures are defined to be 
\[ 
\beta_k (\alpha) = \beta_k \left( \bigcup_{i=1}^{N} B_d(X_i,\alpha) \right)  \quad \text{for} \quad k=0,1,\ldots,d-1. 
\]
Algorithms for computing the Betti numbers are described in the following section.  
Expected values per unit volume, $E \beta_k(\alpha)$, are estimated as mean values calculated from many independent realisations of points in the unit $d$-cube.

\subsection{Betti numbers of alpha shapes}
\label{BettiNumbers}

The union of balls of radius $\alpha$ centred at the points $X_1, \ldots , X_N$ has a geometric dual called the \emph{alpha-shape} that is a subset of the Delaunay triangulation of $X_1, \ldots, X_N$.  
The nerve theorem of topology guarantees that the Betti numbers of the union of balls are identical to those of the dual alpha-shape~\cite{Edels95}.  
Since the alpha-shape is a discrete simplicial complex, the Betti numbers are computable via linear algebra techniques for data in any dimensions~\cite{MunkresAT}. The classical algorithm is impractical for large complexes however, and for points in one, two, or three dimensions there are more effective geometric approaches.  

In 1D there is only the number of connected components to consider, $\beta_0(\alpha)$, and this is determined entirely by distances between adjacent points. 
For points in 2D and 3D, the Betti numbers may be computed via an incremental algorithm due to Delfinado and Edelsbrunner~\cite{DE95} that gives $\beta_k(\alpha)$ at all values of $\alpha$.   The essential aspects of their approach are as follows.

Firstly, the simplices of the Delaunay complex --- the vertices, edges, triangles, and so on --- are ordered by the radius of the smallest sphere that touches the points of the given simplex and contains no other data points.  This radius is called the \emph{alpha-threshold}, $\alpha_T$.  If more than one simplex has the same alpha-threshold, they are ordered from lowest dimension to highest. 
The ordering of simplices $\{\sigma_1, \sigma_2, \ldots, \sigma_n\}$ such that $\alpha_T(\sigma_i) \leq \alpha_T(\sigma_j)$ if $i<j$ is a \emph{filtration}. 
A sequence of subcomplexes $C_j = \bigcup_{i=1}^{j} \sigma_i$ is now built by adding one simplex at a time.  Each $k$-dimensional simplex either creates a new $k$-cycle, or destroys a ($k$--1)-cycle. For example, when an edge is added, it either generates a loop or connects two disjoint components. 
The Betti numbers of $C_{j+1}$ are related to those of $C_j$ by:
\begin{alignat*}{2}
  \beta_k (j+1)  &= \beta_k (j) + 1 \quad  \text{ if $\sigma_{j+1}$ creates a $k$-cycle}  \\
  \beta_{k-1}(j+1)  &= \beta_{k-1} (j) - 1  \text{ if $\sigma_{j+1}$ destroys a ($k$--1)-cycle.}
\end{alignat*}
The problem of determining whether a $k$-simplex creates a $k$-cycle is non-trivial in arbitrary dimension $d$.  
Fast algorithms based on union-find data structures are possible when $k = 1$ and, by Alexander duality~\cite{MunkresAT}, $k = d-1$.  Thus, this incremental approach is effective only for points in $d = 2, 3$ dimensions.  The duality argument requires the complex to be a subset of the $d$-sphere, but this is easily accounted for by adding a point at infinity to the Delaunay complex, and an extra $k$-simplex for each ($k$--1)-face on the convex hull.  

Each $k$-simplex in the filtration, as it is added to the complex, is marked $+1$ if it is found to create a $k$-cycle, and $-1$ if it destroys a ($k$--1)-cycle.  Then the Betti numbers of the alpha-shapes are calculated as 
\begin{alignat}{2}   \label{eq:incr_betti}
  \beta_0 (\alpha)  & = \#\{ +1\text{ vertices}< \alpha\} - \#\{ -1 \text{ edges}< \alpha \}, \\ 
  \beta_1 (\alpha) &  =\#\{ +1\text{ edges}< \alpha\} - \#\{ -1\text{ triangles}< \alpha \}, \nonumber \\ 
  \beta_2 (\alpha) &  =\#\{ +1\text{ triangles}< \alpha\} - \#\{ -1\text{ tetrahedra}< \alpha \}. \nonumber
\end{alignat}
Note that these signature functions may be evaluated at arbitrary fineness in $\alpha$, with no additional computational cost or complexity.  A single traversal of the marked simplices, in the filtration order is all that is required.

\subsection{Periodic boundary conditions}
\label{PeriodicBoundary}

To avoid boundary effects from restricting the domain to the unit $d$-cube, we build the Delaunay complex with periodic boundary conditions.  

For points in 2D, this means the Delaunay complex is a triangulation of the 2-torus, and the Betti numbers for sufficiently large $\alpha$ are $\beta_0 = 1, \beta_1 = 2, \beta_2 = 1$.  The incremental algorithm for computing the Betti numbers is still valid, provided the final triangle added to the filtration is marked as creating a 2-cycle (which we know it must, \emph{a priori}). 
When $\alpha$ is below the percolation threshold, $\beta_1(\alpha)$ may be interpreted as the number of holes per unit area.  If we use periodic boundary conditions, then above the percolation threshold $\beta_1(\alpha)$ includes the cycles around each axis of the torus, and so the number of holes in the unit square, as one would intuitively define them, is really $\beta_1(\alpha) - 2$.   This issue is related to the problem of whether or not to count the spanning cluster when studying the connected components in percolation theory. 

For points in 3D, periodic boundary conditions invalidate the algorithm for the determination of 2-cycles.  The topology of a unit cube with opposite faces identified is that of a 3-torus: $\beta_0 = 1, \beta_1= 3, \beta_2 = 3, \beta_3 = 1$.   Thus, the sequence of subcomplexes obtained from the filtration are subspaces of the 3-torus, not the 3-sphere, and the Alexander duality theorem no longer applies. 

In practice, we apply the duality algorithm to the 2-cycle detection problem anyway.  The result is that exactly three faces (triangles) from the filtration are incorrectly identified as ``destroying 1-cycles'' when in fact they create the three 2-cycles that are homologous to the three coordinate planes of the periodic cube. 
This means there are three extra triangles marked $-1$, and three fewer marked $+1$, than there would be if we had a direct algorithm for detecting 2-cycles. 
All other edges, faces, and tetrahedra are correctly marked $\pm 1$, provided the final tetrahedron is identified as creating a 3-cycle (which we again know \emph{a priori}).
Comparison with the formulas given in (\ref{eq:incr_betti}) shows that for both $k=1,2$,
\[
 \beta_k^{calc} (\alpha) = \beta_k^{true} (\alpha) - \#\{\text{mislabelled triangles}< \alpha \}. 
 \]
If we consider the alpha complex as $\alpha$ increases, it should be clear that the mislabelled triangles have the smallest alpha-thresholds for which each of the three toriodal 2-cycles exist.   This represents the second percolation threshold: a critical radius, $\alpha_2$, above which, with probability one, the unoccupied space no longer percolates. 
Thus, for the mean values of the first and second Betti numbers we have (for $k=1,2$) 
\begin{alignat*}{2}
 E \beta_k^{calc}(\alpha) & = E \beta_k^{true}(\alpha)  & \text{ for } \alpha &< \alpha_2 \\
 E \beta_k^{calc}(\alpha) & = E \beta_k^{true}(\alpha) - 3 \quad & \text{ for }  \alpha &> \alpha_2 .
\end{alignat*}

\subsection{Implementation}

There are three publicly available implementations of alpha-shapes: Edelsbrunner's group \cite{3DAlphaShapes}, Clarkson's hull code \cite{HullCode}, and the CGAL library \cite{cgal:d-as3-06,CGALurl}.  None of these has provision for periodic boundary conditions, and only the first has support for computing the Betti number signatures.  The CGAL library (written in C++) has the most general interface, so we use the CGAL implementation of two- and three-dimensional Delaunay triangulations and alpha shapes and extend it as follows.  

First, $N$ points in the unit square or cube are generated with uniform random coordinates, and each point is given a label. 
Periodic boundary conditions are simulated using translated copies of the original data points with each translated copy of a point given the same label as the original.  
The simplest approach to generating the translated points is to map all the original data points to the 8 adjacent squares in 2D, or the 26 adjacent cubes in 3D.  
This creates a significant overhead in the number of points to be triangulated --- $9N$ and $27N$ respectively.  
There is also an increasing degree of redundancy for large $N$, since almost all of the translated points have no effect on the triangulation within the original domain.    
For reasonable numbers of points ($N > 50$) we can therefore reduce the overhead by translating only the data points in the appropriate half-cube along each axis, leading to  $4N$ and $8N$ points to be triangulated in 2D and 3D respectively. 
The minimum requirement on translating points that guarantees a correct triangulation is in principal even less:  only points that belong to a Delaunay cell whose circumsphere intersects the boundary of the unit square or cube need to be translated to the opposite side \cite{Coveney03}.  
However, we find that the increased complexity of this approach outweighs any saving from the reduced number of translated points.  

The second step is to build the Delaunay complex and alpha shape on the enlarged set of data points using the CGAL library routines.  We must then identify the elements of the Delaunay complex that comprise the periodic domain. The criterion we use is that the centroid of the cell (or face, or edge) is either interior to the unit cube, or lies on one of the $x=0, y=0$, or $z=0$ planes. 
The topological integrity of the Delaunay complex with the periodic boundary conditions is checked via the labels attached to the vertices. 

Finally, we implement a filtration data structure and the incremental Betti number algorithm as described in Section~\ref{BettiNumbers}.   The CGAL alpha shape data structure gives us direct access to the alpha-thresholds of each simplex (i.e., the cells, faces, and edges), so this is relatively straightforward.  The C++ code is available from the author on request. 

The simulations reported in Section~\ref{Results} were performed on a PC with Intel Pentium~4 processor.  The two-dimensional simulations involved 1000 realisations with $\lambda = 10^5$ and ran overnight.  The three-dimensional simulations involved 50 realisations with $\lambda = 10^5$ and took five days.  The dramatic increase in time for the 3D simulations is due to the intrinsic additional complexity of 3D Delaunay complexes and alpha shapes, the extra points needed to simulate periodic boundary conditions, and the need for both a forward and backward traversal of the filtration to mark the simplices.

%%%%%%%%%%%%%%%%%%%%%%%%

\section{Results and analysis}
\label{Results}

In this section, we summarise theoretical results and compare these to data obtained from computer simulations of Poisson point processes in two and three dimensions. 

\begin{figure}
\begin{center}
\includegraphics[width=\columnwidth]{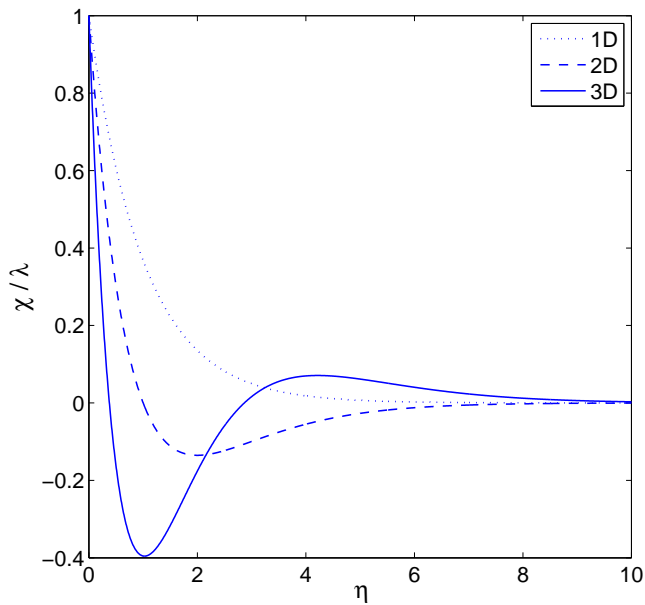}
\caption{Expectations per unit $d$-volume of the Euler characteristic, $\chi / \lambda$, as a function of the reduced density, $\eta = \omega_d \alpha^d \lambda$, for $d = 1,2,3$.  The quantity $\omega_d$ is the $d$-volume of the unit $d$-sphere.}
\label{fig:euler_all}
\end{center}
\end{figure}

\begin{figure}
\begin{center}
\includegraphics[width=\columnwidth]{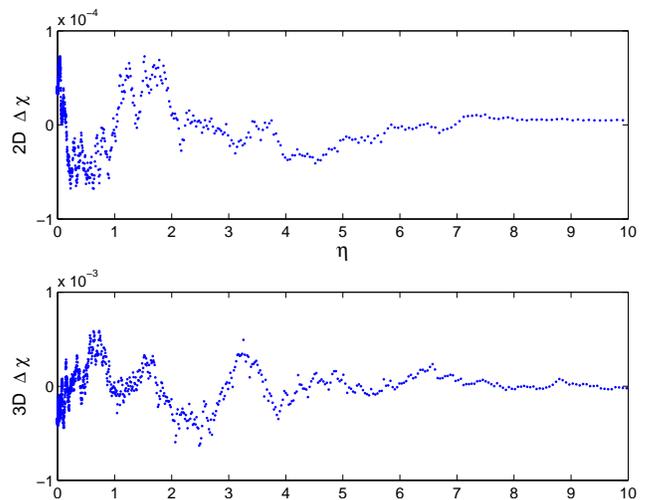}
\caption{Deviation from theory of the computed mean Euler characteristic per unit $d$-volume as a function of the reduced density, $\eta$, for $d = 2,3$.  In the 2D case, the mean Euler characteristic is computed from 1000 simulations of a Poisson point process with $\lambda = 10^5$ in the unit square with periodic boundary conditions.  In the 3D case, we have used 50 simulations with $\lambda =10^5$ in the unit cube with periodic boundary conditions.  }
\label{fig:delta_chi_2D_3D}
\end{center}
\end{figure}

\subsection{1D}

The expected number of components per unit length in a 1D Boolean model is well known \cite{HallBook}.  If the intensity of the Poisson-point process is $\lambda$, and the shapes are line segments of length $2\alpha$ then
\[  E \chi (\alpha) =  E \beta_0 (\alpha) = \lambda e^{-2\alpha\lambda} . \]
This result is included for completeness and ease of comparison with the two- and three-dimensional cases, see Figure~\ref{fig:euler_all}.

\subsection{2D}

\begin{figure}
\begin{center}
\includegraphics[width=\columnwidth]{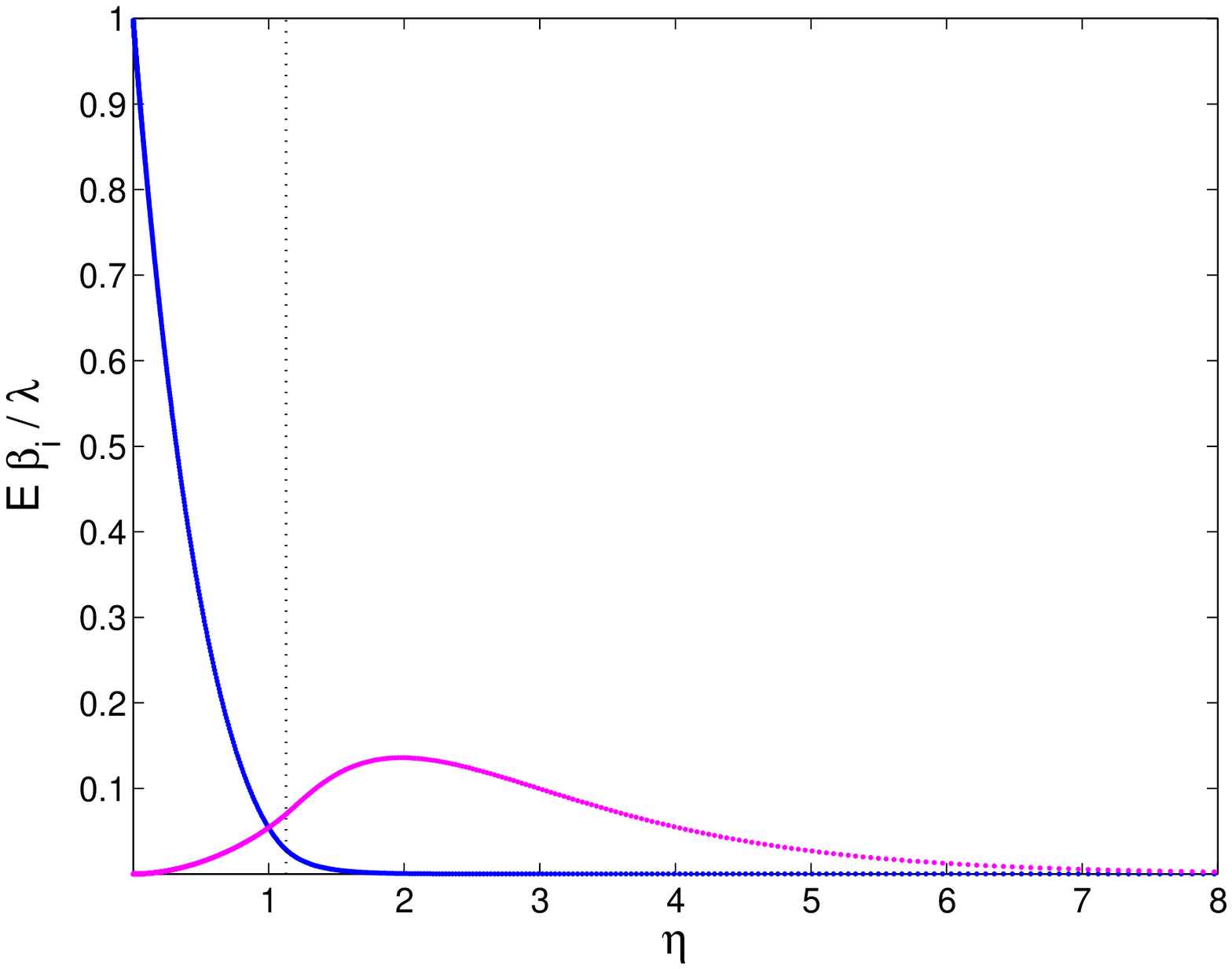} \\
\vspace{-6.6cm} \hspace{2.7cm}
\includegraphics[width=0.6\columnwidth]{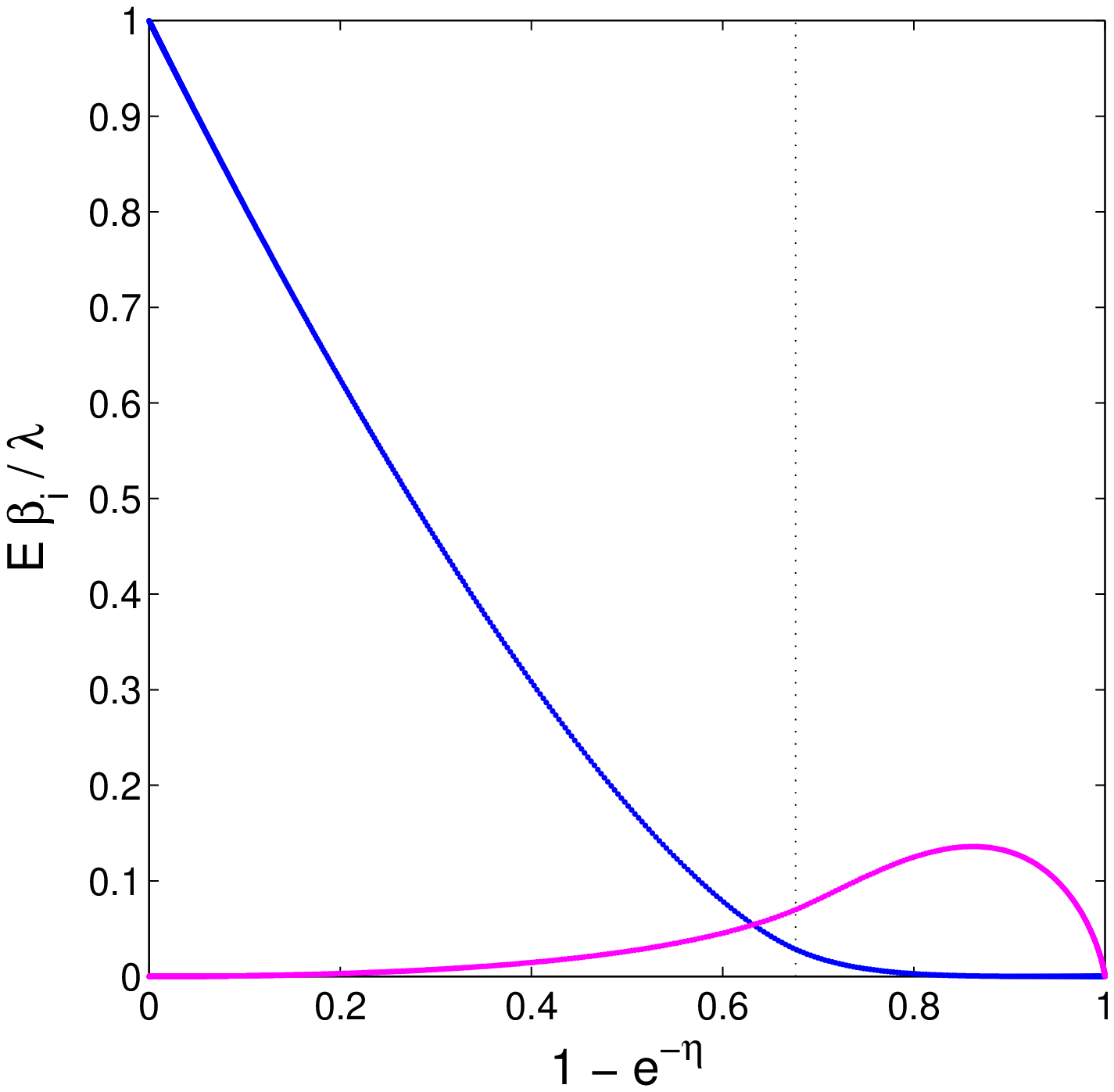}\\
\vspace{1.5cm}
\caption{2D Betti numbers.  Results from 1000 simulations of, on average, $10^5$ points in the unit square.  Mean values of the Betti numbers per unit area, $\beta_0$ (blue dots) and $\beta_1$ (magenta dots) are given as  functions of the reduced density, $\eta$ (main) and the disc area fraction $\phi = 1-e^{-\eta}$ (inset).  The percolation threshold is marked by the dotted vertical line at $\eta_c = -1.1280586$, or equivalently $\phi_c = 0.676339$. }
\label{fig:2D_meanbetti}
\end{center}
\end{figure}

\begin{figure}
\begin{center}
\includegraphics[width=\columnwidth]{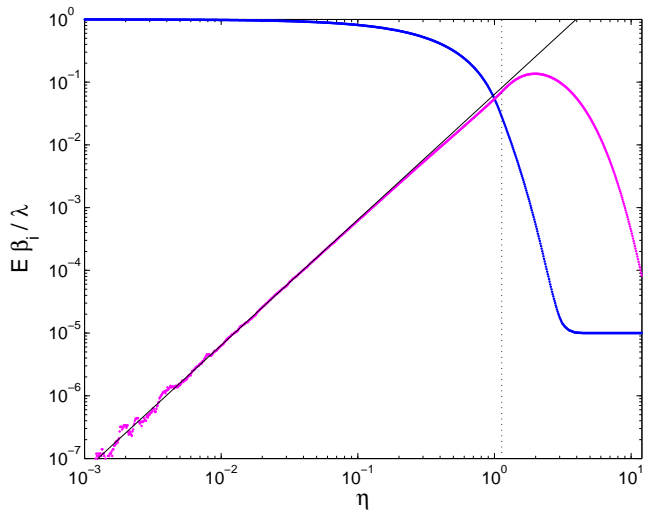}
\caption{2D Betti numbers.  Exactly the same data as in Fig.~\ref{fig:2D_meanbetti} but plotted here with logarithmic axes to emphasise the quadratic scaling of $\beta_1$ at small $\eta$.  The solid black line shows the leading order behaviour $E \beta_1 / \lambda \sim 0.0640 \eta^2$ derived in Section~\ref{PDC_2D}.}
\label{fig:2D_meanbetti_loglog}
\end{center}
\end{figure}

For discs of radius $\alpha$ centred at points from a 2D Poisson point process of intensity $\lambda$, we study the expectation per unit area of the following topological quantities:
the number of components, $\beta_0(\alpha)$,  the number of independent cycles, $\beta_1(\alpha)$, and the Euler characteristic, $\chi = \beta_0 - \beta_1$.
For our simulations, we use an intensity of $\lambda = 10^5$ in the unit square, and compute mean values of the Betti numbers from 1000 realisations.  Results are presented in Figures \ref{fig:2D_meanbetti} and~\ref{fig:2D_meanbetti_loglog}. 
In the plots of these figures we mark the 2D continuum percolation threshold from  \cite{QuintTorqZiff00} of $\eta_c = 1.1280586$.  The critical value is included as a reference point only, since the Betti numbers are not sensitive indicators of percolation.  

The expectation per unit area of the Euler characteristic is known from stochastic geometry to be \cite{HallBook,StoyanBook}
\begin{equation} \label{eq:2Deuler}
\begin{split}
E \chi (\alpha) & = \lambda (1 - \pi \alpha^2\lambda ) e^{-\pi \alpha^2\lambda}  \\
	& =  \lambda (1 - \eta) e^{-\eta} . 
\end{split}
\end{equation}
This expression is more naturally a function of the reduced density, $\eta = \pi \alpha^2 \lambda$, and we often use $\eta$ as the independent variable rather than the radius $\alpha$.   
The differences between the expression (\ref{eq:2Deuler}) and the computed mean values of the Euler characteristic are less than $10^{-4}$, and decrease as $\eta$ increases, see Fig.~\ref{fig:delta_chi_2D_3D}. 

\begin{figure}
\begin{center}
\includegraphics[width=\columnwidth]{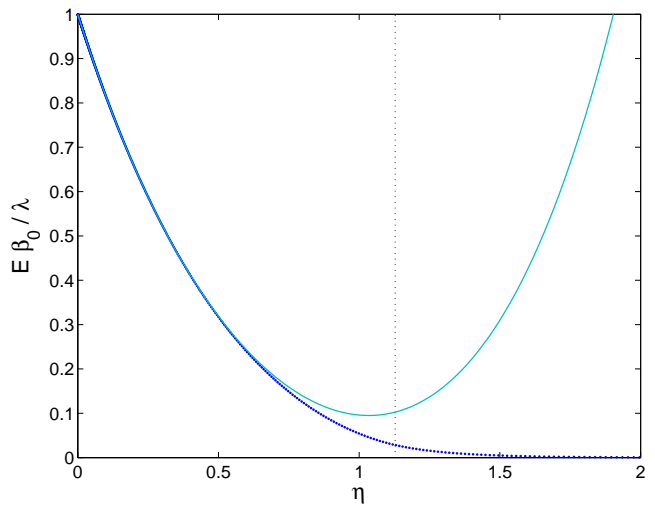}
\caption{2D connected components. Here we compare the computed mean values (blue dots) of the number of connected components, $\beta_0 / \lambda$ with the theoretical expansion (pale blue line) for small $\eta$ given in (\ref{eq:2D_b0_expan}).}
\label{fig:2D_mnb0_expansion}
\end{center}
\end{figure}

The connected components of randomly distributed overlapping discs are studied extensively in percolation theory.  It is common in this context to express the expected total number of components per unit area as the sum 
\[ E \beta_0 = \sum_{k=0}^{\infty} \rho_k \]
where $\rho_k$ is the expected number of $k$-mers per unit area (a $k$-mer is a cluster built from $k$ discs). 
Although there are no known analytic expressions for $E \beta_0$ as a function of disc radius $\alpha$, integral expressions for $\rho_k$ and low-density expansions are given in \cite{QuintanillaTorquato96}.
The expansions for $\rho_k$ are given in terms of the reduced density for the limit $\eta \to 0$ and presented in Table~\ref{tab:expansions_2d} for reference. 
Their sum gives 
\begin{equation}\label{eq:2D_b0_expan}
\begin{split} 
 E \beta_0 (\eta) / \lambda =  1 - 2\eta  & +1.5641 \eta^2 - 0.6878 \eta^3 \\ 
      & + 0.2197\eta^4 + O(\eta^5) . 
\end{split}
\end{equation}
A comparison between this expansion and the computed mean values obtained from simulations is shown in Fig.~\ref{fig:2D_mnb0_expansion}.  There is extremely close agreement for $\eta < 0.5$. 

\begin{table}
\caption{Coefficients in the expansions of $\rho_k$ for the 2D Poisson-Boolean model of discs with radius $\alpha$ for the limit $\eta = \pi \lambda \alpha^2 \to 0$. Results are from \cite{QuintanillaTorquato96}. }
\begin{center}
\begin{tabular}{cccccc}
\hline 
  &  $\eta^0$  &  $\eta^1$  &  $\eta^2$   &  $\eta^3$  & $\eta^4$ \\
\hline 
$\rho_1 / \lambda $
  & 1  & -4  &  8   &  -10.6667  &  10.6667 \\
$ \rho_2 / \lambda $
  &     &  2  & -11.3079  &  32.2915  &  -62.0415 \\
$ \rho_3 / \lambda $
  &    &     &  4.8720  &  -35.3346  &  129.6895 \\
$ \rho_4 / \lambda $
  &    &    &    &  13.022   & -114.823 \\
$ \rho_5 / \lambda $
  &  &  &  &  & 36.728 \\
\hline 
\end{tabular}
\end{center}
\label{tab:expansions_2d}
\end{table}

An expansion for $E \beta_1$ may be deduced from the expressions for $E \beta_0$ and $E \chi$  above.  However, we make an independent analysis of the shape of Poisson-Delaunay cells in Section~\ref{PDC_2D} and find that for small $\eta$ 
\begin{equation}
	E \beta_1 (\eta) / \lambda = 0.0640 \eta ^2  + O(\eta^3).
\end{equation}
Our simulation data show that this leading order behaviour holds for $\eta < 0.3$, see Fig.~\ref{fig:2D_meanbetti_loglog}. 

The logarithmic axes used in Fig.~\ref{fig:2D_meanbetti_loglog} show that $E\beta_0/\lambda$ levels out at $10^{-5}$. This is exactly as expected since for large radius, the alpha-shape has one connected component, and the value of $\lambda$ is $10^5$ for these simulations.  
As discussed in Section~\ref{PeriodicBoundary}, periodic boundary conditions mean that for sufficiently large radius we know $\beta_1 = 2$.  
Thus we would expect to see $E\beta_1/\lambda$ level out at $2*10^{-5}$ in  Fig.~\ref{fig:2D_meanbetti_loglog}, but the range in this plot does not extend to large enough $\eta$.  
This shows that periodic boundary effects are negligible for the data from these simulations. 

\subsection{3D}

\begin{figure}
\begin{center}
\includegraphics[width=\columnwidth]{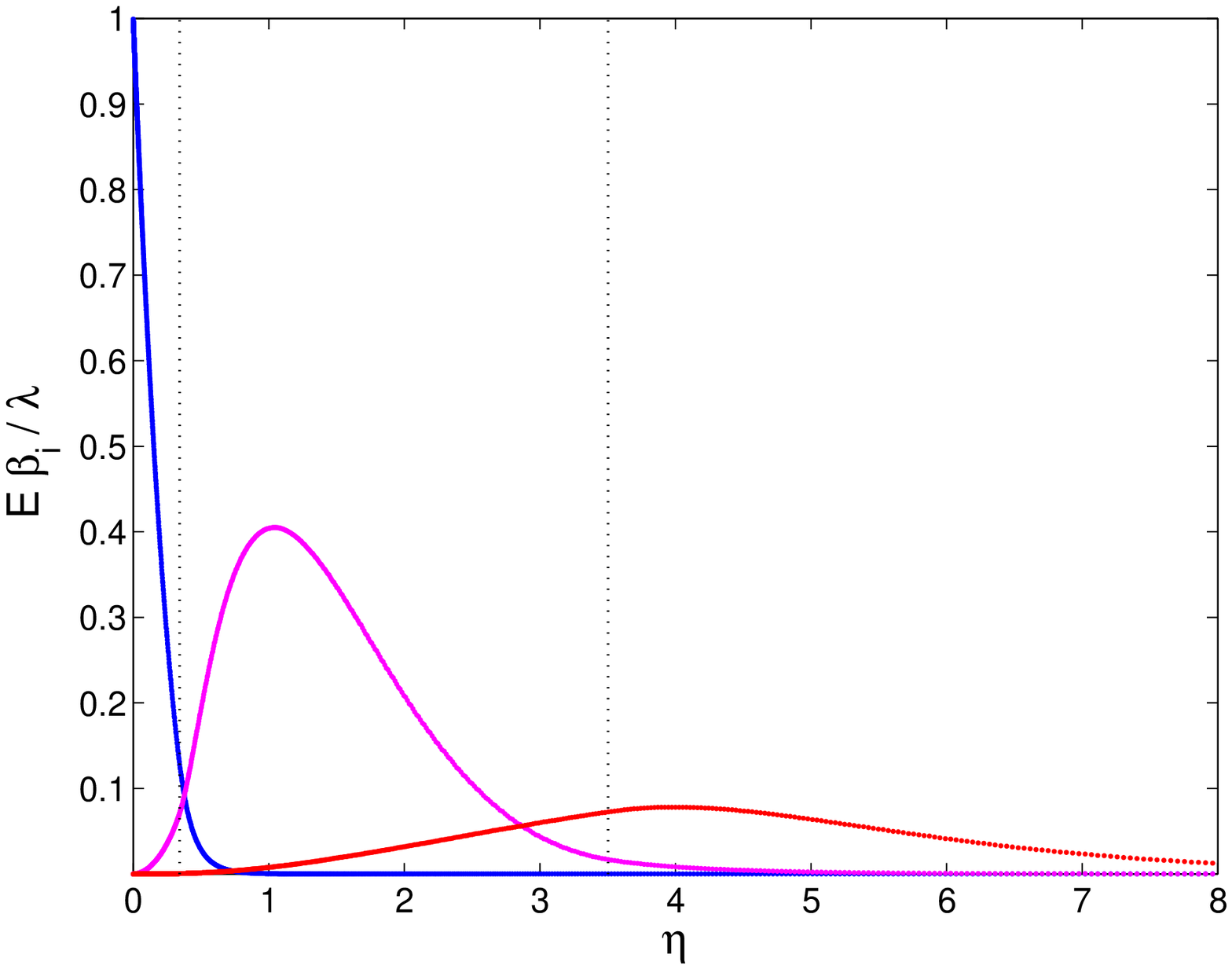} \\
\vspace{-6.6cm} \hspace{2.7cm}
\includegraphics[width=0.6\columnwidth]{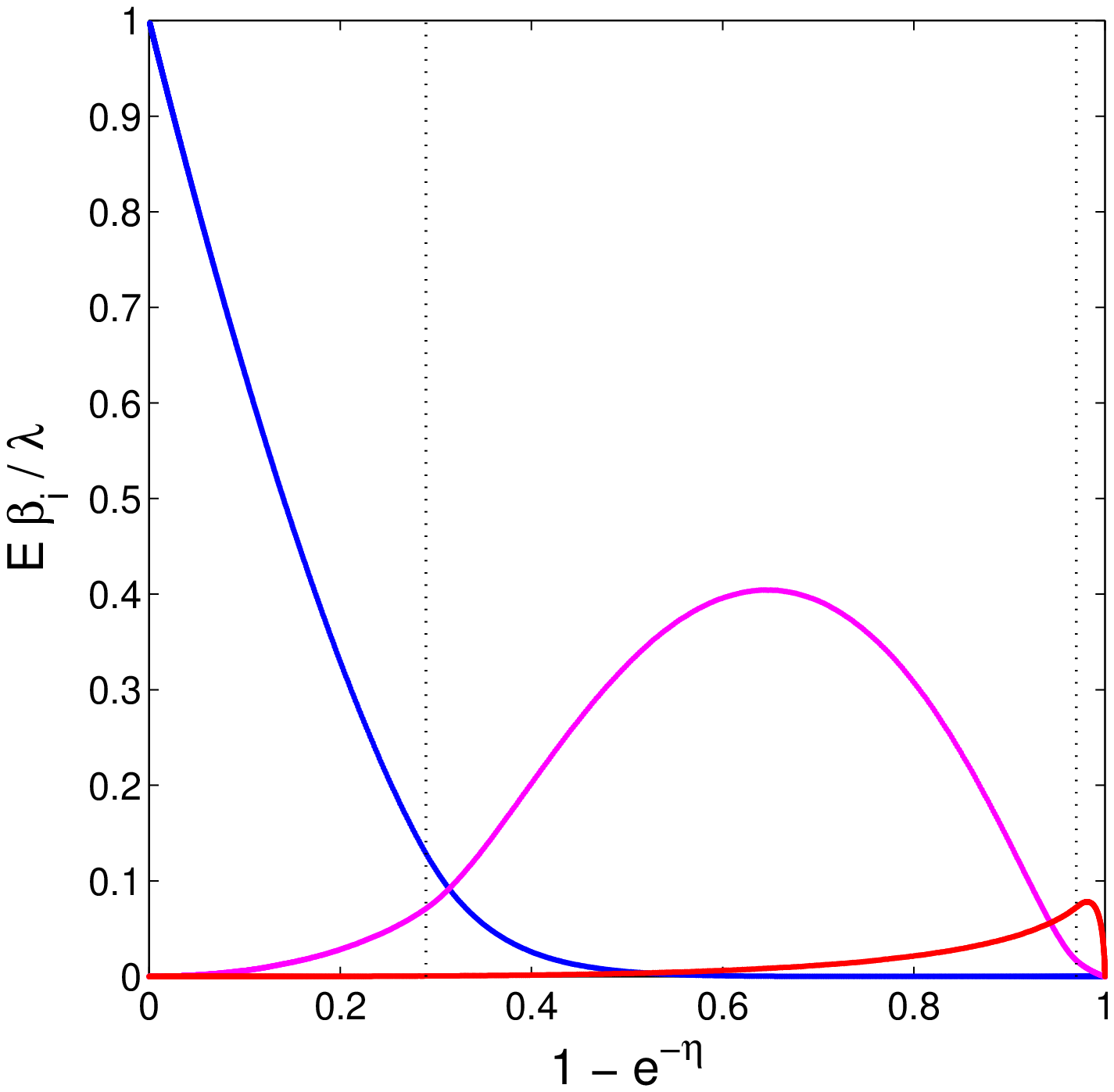}\\
\vspace{1.5cm}
\caption{3D Betti numbers.  Results from 50 simulations of, on average, $10^5$ points in the unit cube.  Mean values of the Betti number per unit volume, $\beta_0$ (blue dots), $\beta_1$ (magenta dots), and $\beta_2$ (red dots), are plotted as functions of the reduced density $\eta = \tfrac{4}{3}\pi \alpha^3$ (main) and ball volume fraction $\phi = 1-e^{-\eta}$ (inset).  The two critical densities from percolation theory are marked by dotted black lines at $\eta_1 = 0.341889$  ($\phi_1 = 0.289573$) and  $\eta_2 = 3.5032$ ($\phi_2 = 0.9699$). } 
\label{fig:3D_meanbetti}
\end{center}
\end{figure}

\begin{figure}
\begin{center}
\includegraphics[width=\columnwidth]{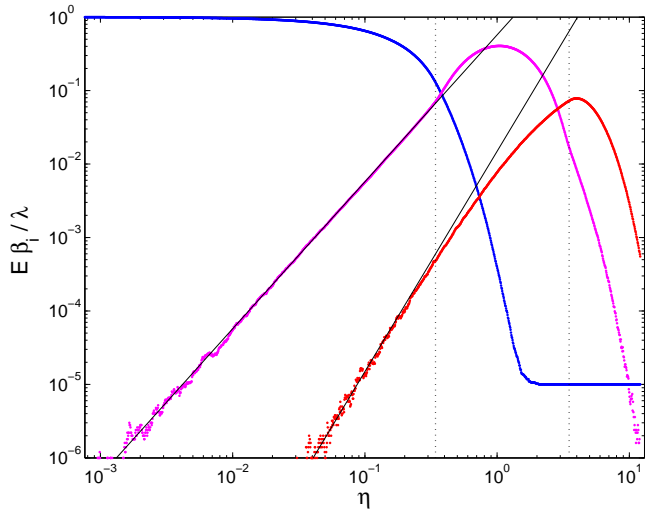}
\caption{3D Betti numbers.  The same results as in Fig.~\ref{fig:3D_meanbetti} but plotted with logarithmic axes to show the power-law scaling of $\beta_1$ and $\beta_2$ for small $\eta$. 
The solid black lines show the leading order behaviour of $E\beta_1/\lambda \sim 0.5747\eta^2$, and $E \beta_2/\lambda \sim 0.015 \eta^3$ derived in Sections~\ref{PDC_3D1} and~\ref{PDC_3D2}. }
\label{fig:3D_meanbetti_loglog}
\end{center}
\end{figure}

In the three-dimensional Poisson-Boolean model of balls with radius $\alpha$, the relevant topological quantities are the number of components, $\beta_0(\alpha)$, the number of independent handles, $\beta_1(\alpha)$, the number of enclosed voids, $\beta_2(\alpha)$, and the Euler characteristic,  $\chi = \beta_0 - \beta_1 + \beta_2$.
For the simulations we use an intensity of $\lambda = 10^5$ in the unit cube and compute mean Betti numbers from 50 realisations.  Results are presented in Figures~\ref{fig:3D_meanbetti} and~\ref{fig:3D_meanbetti_loglog}.  We again mark the continuum percolation thresholds in these plots as reference points.  Recall that in three-dimensional percolation there are two critical densities: $\eta_1 = 0.341889$ \cite{LorenzZiff01}, is the point above which a spanning cluster exists with probability one, and $\eta_2 = 3.5032$ \cite{Rintoul00} is the density above which the unfilled space no longer percolates. 

The expectation per unit volume of the Euler characteristic is again known from stochastic geometry \cite{StoyanBook} to be:
\begin{equation}
E \chi (\eta)  = \lambda (1 - 3\eta + \frac{3\pi^2}{32}\eta^2 ) e^{-\eta} , 
\end{equation}
where $\eta$ is the reduced density $\eta = \tfrac{4}{3}\pi \lambda \alpha^3$.  Our computed mean values match this expression closely, with differences less than $10^{-3}$ and decreasing with $\eta$ as shown in Fig.~\ref{fig:delta_chi_2D_3D}. 

\begin{figure}
\begin{center}
\includegraphics[width=\columnwidth]{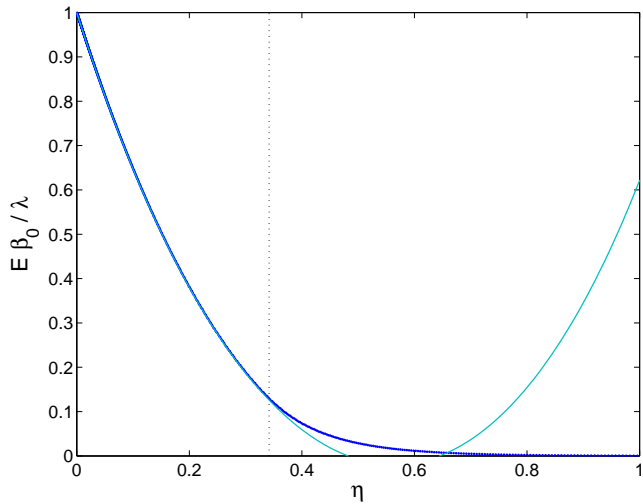}
\caption{3D connected components.  The computed mean values of $\beta_0/ \lambda$ (blue dots) compared to the low-intensity expansion (pale blue line) given in (\ref{eq:3D_b0_expan}).}
\label{fig:3D_mnb0_expansion}
\end{center}
\end{figure}

As for the 2D model, the expected total number of components per unit volume may be expressed as the sum of numbers of $k$-mers.  
Integral expressions and low-density expansions for the expected number of $k$-mers per unit volume, $\rho_k$, are given in \cite{QuintanillaTorquato96} and repeated here in Table~\ref{tab:expansions_3d}. 
From these expansions we find that for $\eta \to 0$
\begin{equation}\label{eq:3D_b0_expan} 
\begin{split}
  E \beta_0 (\eta) / \lambda  =  1 - 4 \eta & + 5 \eta^2 - 2.7431\, \eta^3 \\
  	& + 1.3646 \, \eta^4 + O(\eta^5) .
\end{split}
\end{equation}
The computed mean values match this expansion extremely closely for $\eta < 0.3$, see Fig.~\ref{fig:3D_mnb0_expansion}. 

\begin{table}
\caption{Coefficients in the expansions of $\rho_k$ for the 3D Poisson-Boolean model of balls with radius $\alpha$ for the limit $\eta = \tfrac{4}{3} \pi \lambda \alpha^3 \to 0$. Results are from \cite{QuintanillaTorquato96}. }
\begin{center}
\begin{tabular}{cccccc}
\hline 
  &  $\eta^0$  &  $\eta^1$  &  $\eta^2$   &  $\eta^3$  & $\eta^4$ \\
\hline 
$\rho_1 / \lambda $
  & 1  & -8  &  32   &  -85.3333  & 170.6667 \\
$ \rho_2 / \lambda $
  &     &  4  & -49  & 302.2238  &  -1250.5030 \\
$ \rho_3 / \lambda $
  &    &     &  22  &  -359.4203  &  2959.1209 \\
$ \rho_4 / \lambda $
  &    &    &    &  139.7867   & -2842.60 \\
$ \rho_5 / \lambda $
  &  &  &  &  & 964.68 \\
\hline 
\end{tabular}
\end{center}
\label{tab:expansions_3d}
\end{table}

The leading order behaviour for $\beta_1$ and $\beta_2$ is derived from the Poisson-Delaunay analysis in Sections~\ref{PDC_3D1} and \ref{PDC_3D2} where we show  that for small $\eta$
\begin{align} \label{eq:3D_low_dens}
 E \beta_1 (\eta) / \lambda  & = 0.5747\, \eta ^2  + O(\eta^3)  \\
 E \beta_2(\eta) / \lambda & = 0.015\, \eta^3 + O(\eta^4). 
\end{align}
Again, the computed mean values show exactly this leading order behaviour for $\eta < 0.3$, see Fig.~\ref{fig:3D_meanbetti_loglog}. 

Recall from Section~\ref{PeriodicBoundary} that with periodic boundary conditions and $\eta > \eta_2$, the second percolation threshold, there is a systematic error in the computed mean values of $\beta_1$ and $\beta_2$.  For the data presented here, the error is $3*10^{-5}$, which is three orders of magnitude less than the value of $E\beta_1(\eta_2)/\lambda$ and four orders less than $E\beta_2(\eta_2)/\lambda$.  
A close inspection of Fig.~\ref{fig:3D_meanbetti_loglog} shows that this error is significant only for the  computed values of $E \beta_1(\eta)/\lambda$ with $\eta > 6$.

%%%%%%%%%%%%%%%%%%%%%%%%%%%%%%%%%%%
\section{Poisson-Delaunay Cell analysis of alpha shapes}
\label{Theory}

The probability distribution for the size and shape of a cell in the Delaunay complex of a Poisson point process is completely characterised by a result due to Miles~\cite{Miles74} and given in  Eq.(\ref{eq:PDCpdf}).  
The criteria for a simplex from the Poisson-Delaunay complex to belong to an alpha shape are based only on the size and shape of that simplex, and that of its adjacent simplices. 
The ergodicity of the Poisson-Delaunay complex means that the expected number of $k$-dimensional simplices, $\sigma$, in a bounded region, $R$, that satisfy condition $A$, is related to the probability that a randomly selected simplex has property $A$:
\[
	E \# \{ \sigma \in R \; | \; \sigma \text{ is } A \} = \lambda_k ||R|| Pr(A) , 
\]
where $\lambda_k$ is the intensity of the $k$-dimensional cells, not the vertices (which have intensity $\lambda_0 = \lambda$). 
Since the Betti numbers of alpha-shapes are determined by numbers of simplices with certain properties, see (\ref{eq:incr_betti}), the Poisson-Delaunay cell (PDC) distribution can be used to obtain results about the Betti numbers of an alpha-shape.   
 
This section summarises the relevant results about the PDC distributions in two- and three-dimensions, and then derives low-intensity expansions for the expectation per unit area of $\beta_1$ in 2D and expectation per unit volume of $\beta_1$ and $\beta_2$ in 3D. 

\subsection{Distributional properties of PDCs}

For an extensive review of Poisson Delaunay cells, see the book \emph{Spatial Tesselations} (2nd ed., Section 5.11)\cite{OkabeBook}. 

We start by considering a Poisson point process with intensity 
$\lambda$ in $\mathbb{R}^m$.
A Poisson Delaunay cell is an $m$-dimensional 
simplex, i.e., the convex hull of $m+1$ points $X_0,\ldots,X_m$ from 
the Poisson point process, such that there exists an $(m$--1)-sphere that has each 
point $X_0,\ldots,X_m$ on its boundary and no other points either on its boundary or in its interior.  This is the circumsphere of $X_0,\ldots,X_m$; let $\mathbf{c}$ and $r$ denote the circumcenter and circumradius respectively.
The vertices of a PDC are given in vector form by 
$\mathbf{X}_i = \mathbf{c} + r \mathbf{u}_i$, 
where $\mathbf{u}_i$ is a unit vector pointing from the centre to 
the point $X_i$.  
The circumradius is a measure of the size of a simplex, and the unit 
vectors $\mathbf{u}_{i}$ specify its shape. 

The distribution of PDCs is completely specified by the following 
probability density function (pdf). 
This result is due to Miles~\cite{Miles74}, and implies that the 
circumradius of a PDC is independent of the positions of its vertices;  
\begin{equation}
\label{eq:PDCpdf}
h_m(r,\mathbf{u}_0,\ldots, \mathbf{u}_{m}) 
   = a(\lambda,m) \Delta_{m}r^{m^2-1} \exp(-\lambda \omega_{m} r^{m}).
\end{equation}
The constant $\omega_m = \pi^{m/2}/\Gamma(m/2 + 1)$ is the volume of the 
$m$-dimensional unit sphere, and 
\[
  a(\lambda,m) = \frac{\pi^{(m^{2}+1)/2}\Gamma(m^{2}/2)
                       \{2\lambda\Gamma[(m+1)/2]\}^{m}}
		      {m^{m-2}\Gamma(m/2)^{2m+1}\Gamma[(m^{2}+1)/2]}. 
\]
The dependence of $h_m$ on the $\mathbf{u}_{i}$ is hidden in the function 
$\Delta_{m}$, defined as the volume of the $m$-simplex with 
vertices at $\mathbf{u}_0,\ldots,\mathbf{u}_{m}$.  
It is therefore a constant, $1/(m!)$, times the determinant of a square 
matrix with $m$ rows containing the vectors 
$(\mathbf{u}_{i}-\mathbf{u}_{0})$, for $i = 1,\ldots,m$.

Various distributional properties of PDCs can be derived from this pdf. In particular, Muche~\cite{Muche96,Muche98} has simplified the pdf for the two- and three-dimensional cases, finding in 2D:
\begin{equation}\label{eq:PDC_2Dpdf}
 \begin{split}
   f_2(r, \phi_1, \phi_2) {=}& 2 (\pi \lambda)^2 r^3 \exp(-\lambda \pi r^2) \,\mathbf{\cdot} \\
         {}& \frac{2}{3\pi} \sin\frac{\phi_1}{2} \sin\frac{\phi_2}{2} \sin\frac{\phi_1+\phi_2}{2}, \\
 \end{split}
\end{equation}
where $0\leq r <\infty$ is the circumradius, $0 \leq \phi_2 \leq 2\pi-\phi_1$, and $0 \leq \phi_1 < 2\pi$ are the central angles $X_0 c X_1$ and $X_1 c X_2$. 

In 3D, we can choose a coordinate system so that the circumcentre is at the origin, and three points $(X_1, X_2, X_3)$ of the tetrahedron lie in the plane $x = \cos \theta$, where $\theta$ is the angle between the normal to this face (i.e., the positive $x$-axis) and one of its vertices.  The $y$- and $z$-coordinates of the vertices in this triangular face are then determined by the central angles $\phi_1$ and $\phi_2$.   
The distributional properties of this face are those of a ``typical'' face in a Poisson-Delaunay complex.
The fourth vertex of the tetrahedron is specified by the height of the tetrahedron, $h$, and an angle $\gamma$.  Muche~\cite{Muche96} showed that the pdf for a Delaunay tetrahedron separates into factors: 
\begin{equation}\label{eq:PDC_3Dpdf}
  f_3(r,\theta,h,\phi_1,\phi_2,\gamma) = 
	f_R(r) f_{\Theta,H}(\theta,h) f_{\Phi}(\phi_1,\phi_2) f_{\Gamma}(\gamma) 
\end{equation}
with marginal densities: 
\begin{alignat*}{2}
	f_R (r) &= \frac{32\pi^3\lambda^3}{9} r^8 \exp(-\frac{4\pi\lambda}{3}r^3), \; 0 \leq r < \infty,\\
	f_{\Theta,H}(\theta,h) &= \frac{105}{64} \, h \sin^5\theta,  \\
		& \qquad 0\leq h < 1+\cos\theta,  \quad 0\leq \theta < \pi, \nonumber \\ 
	f_{\Phi}(\phi_1,\phi_2) &= \frac{16}{3\pi^2} \left( \sin\frac{\phi_1}{2} \sin\frac{\phi_2}{2} \sin\frac{\phi_1+\phi_2}{2} \right)^2,  \\
		& \qquad 0 \leq \phi_2 \leq 2\pi-\phi_1, \quad 0\leq \phi_1 < 2\pi, \nonumber  \\ 
	f_{\Gamma}(\gamma) &= \frac{1}{2\pi},   \qquad 0 \leq \gamma < 2\pi . 
\end{alignat*}

\subsection{Empty triangles in 2D}
\label{PDC_2D}

We now derive conditions on a Poisson Delaunay cell in $\mathbb{R}^{2}$
that guarantee the 2-simplex is excluded from the alpha-shape, but all its edges are included in the alpha-shape. 
This implies the existence of a non-bounding 1-cycle (a hole)   
that we refer to as a $\tri$-loop. 
From (\ref{eq:PDC_2Dpdf}) we can write down an integral for 
the probability, $P_{\tri}(\alpha)$, that a PDC gives us a $\tri$-loop in the alpha-shape. 
Then, in a region $R$, the expected total number of holes at any radius $\alpha$, is bounded by 
\[
	E \beta_1(\alpha) \geq \lambda_2 ||R|| P_{\tri}(\alpha),  
\]
where $\lambda_2 = 2\lambda$ is the intensity of triangles in a 2D Poisson-Delaunay complex. 

The conditions on the size and shape of a triangle to generate $\tri$-loop are that
\begin{enumerate}
 \item all edges belong to the alpha-shape, i.e. $l_{\max} < 2\alpha$;
 \item the 2-simplex is excluded from the alpha-shape, i.e. the circumradius satisfies $r > \alpha$;
 \item the circumcenter must be interior to the triangle, i.e. it is an acute triangle and the largest vertex angle satisfies $\phi_{\max} < \pi/2$. 
\end{enumerate}

The length of an edge in a triangle is related to the angle at the opposite vertex via 
$l= 2 r\sin\phi$. Thus, the condition $l_{\max} < 2 \alpha$ implies that 
$r < \alpha/\sin(\phi_{\max})$. 
The marginal density for the largest angle at a vertex of a Poisson-Delaunay triangle is known to be~\cite[p.398]{OkabeBook}
\[
  f_{\max}(\phi) = 
  \begin{cases} 	
    \begin{aligned}
     \tfrac{2}{\pi}[(3\phi & - \pi)\sin2\phi  \\
     	&- \cos 2\phi + \cos 4\phi], 
    \end{aligned}  
         &  \tfrac{\pi}{3} \leq \phi < \tfrac{\pi}{2}  \\
     \tfrac{4}{\pi}[ \sin\phi + (\pi - \phi)\cos\phi ]\sin\phi, 
        & \tfrac{\pi}{2} \leq \phi < \pi  .
   \end{cases}
 \]
Thus an integral expression for $P_{\tri}(\alpha)$ is
\begin{equation} 
\label{eq:PtriIntegral}
    P_{\tri} = \int_{\frac{\pi}{3}}^{\frac{\pi}{2}} \!\!
               \int_{\alpha}^{\alpha/\sin\phi} 
	         2(\pi\lambda)^2 r^3 e^{-\pi\lambda r^2} f_{\max}(\phi) dr \, d\phi .
\end{equation}

To evaluate this integral, we start with an expression for the indefinite integral of the 
circumradius pdf:
\[
    G(r) = \int  2(\pi\lambda)^{2} r^{3} e^{-\pi \lambda r^{2}} dr 
	 = -(\pi\lambda r^{2} + 1) e^{-\pi\lambda r^{2}}.
\]
Evaluating with the limits of integration from (\ref{eq:PtriIntegral}) 
we obtain
\[
     G(\eta,\phi) = 
       (\eta + 1)e^{-\eta} - 
	\left( \frac{\eta}{\sin^2\phi} + 1 \right)
          e^{-\eta/\sin^2\phi} 
\]
where we have simplified notation by using the reduced density,
$\eta = \pi\lambda \alpha^{2}$. 
The second integral with respect to the angle $\phi$ does not have an analytic solution. 
However, we can obtain an approximate expression for small $\alpha$ (i.e. small $\eta$) by using a Taylor expansion.  First note that 
\[
    (x + 1) e^{-x} = 1 - \frac{x^2}{2} + \frac{x^3}{3} - \frac{x^4}{8} + \cdots 
\]
The first term in the series for $G(\eta,\phi)$, and
consequently the leading order term of $P_{\tri}$, is therefore 
 $\eta^2$.  
The coefficient of $\eta^j$, for $j\geq2$ in the series for 
$P_{\tri}$ is therefore given by the integral
\[
  P_{\tri}^{(j)} = \frac{(-1)^{j-1}(j-1)}{j!}
           \int_{\frac{\pi}{3}}^{\frac{\pi}{2}} 
           \left( 1 - \frac{1}{\sin^{2j}\phi} \right) 
	    f_{\max}(\phi) d\phi \, . 
\]
To evaluate these integrals requires only standard techniques 
from real calculus; for the first few terms we have:
\[
   P_{\tri} = 0.03200 \eta^{2} - 0.03422 \eta^{3} + 0.01835 \eta^{4} + O(\eta^5) .
\]
Since the intensity of Delaunay cells is $2\lambda$, we have that the expectation per unit area of the first Betti number for small $\eta$ is 
\[
 E \beta_1(\eta)  \geq 2\lambda P_{\tri} \sim 0.0640 \lambda \eta^2.
 \]
In fact this lower bound on $E \beta_1(\eta)$ is an asymptotic expression as $\eta \to 0$.  
This is because a connected cluster of at least three discs is needed to create a $\tri$-loop, and at least four overlapping discs are necessary to create a non-bounding 1-cycle with four or more edges.  We know from the cluster expansions in Table~\ref{tab:expansions_2d}, however,  that the leading order term as $\eta \to 0$ for the number of $k$-mers is $\eta^{k-1}$.  Thus there can be no other contribution to the $\eta^2$ coefficient in a series expansion of $E\beta_1(\eta)$. 
Indeed, in the limit of small $\eta$, our simulations show exactly this behaviour --- see Fig.~\ref{fig:2D_meanbetti_loglog}.

\subsection{Empty triangles in 3D}
\label{PDC_3D1}

We can derive a similar integral expression to that above for the probability of a $\tri$-loop in $\mathbb{R}^{3}$.  However, in three dimensions not every $\tri$-loop represents an independent 1-cycle in the homology group.  To see why this is the case, consider a cage consisting of the six edges of a tetrahedron.  There are four $\tri$-loops in this cage but only three independent 1-cycles, since the fourth $\tri$-loop is the sum of the other three. 
Nevertheless, using a similar argument to that in the previous section, in the limit of small $\alpha$, or small $\eta = \frac{4}{3}\pi\lambda\alpha^3$, we can assume that the $\tri$-loops are isolated and that in a region $R$,  
\[
	E \beta_1(\alpha) \sim \lambda_2 ||R|| P_{\tri}(\alpha)
\]
where $\lambda_2 = \frac{48}{35} \pi^2 \lambda$ is the intensity of faces in a 3D Poisson-Delaunay complex. 

The conditions for the existence of a $\tri$-loop in a 3D Poisson-Delaunay complex are essentially the same as those in two dimensions, except that they now apply to a typical face of a 3D PDC:
\begin{enumerate}
 \item all edges of the typical face belong to the alpha-shape, i.e. $l_{\max} < 2\alpha$;
 \item the circumradius of the face satisfies $\rho > \alpha$;
 \item the circumcenter of the face must be in the relative interior of the triangle, i.e., the largest vertex angle in a typical face satisfies $\phi_{\max} < \pi/2$. 
\end{enumerate}
We use the relationship between edge-length and opposite angle again so that condition~1 above becomes $\rho < \alpha/\sin(\phi_{\max})$.
We also use the relationship between the face circumradius ($\rho$) and tetrahedron circumradius ($r$) of  
$\rho = r \sin\theta$. 
Thus, an integral expression for $P_{\tri}(\alpha)$ in the three-dimensional setting is:
\begin{equation*}
  P_{\tri} = \int_{\frac{\pi}{3}}^{\frac{\pi}{2}} \!\!\!
               \int_{\alpha}^{\alpha/\sin\phi} \!\!\!\!
               \int_{0}^{\pi} \!\!
	        f_{\max}(\phi) f_R\!\left(\tfrac{\rho}{\sin\theta}\right) \tfrac{1}{\sin\theta} f_{\Theta}(\theta)  d\theta \, d\rho \, d\phi .
\end{equation*}
The densities are
\[
 f_{\Theta}(\theta) = \frac{105}{64} \int_{0}^{1+\cos\theta}  h \sin^5\theta dh 
 		=  \frac{105}{128}\sin^5\theta(1+\cos\theta)^2 ,
\]
\[
 f_R\left(\frac{\rho}{\sin\theta}\right)  = \frac{3}{2}(\tfrac{4}{3}\pi\lambda)^{3} \frac{\rho^{8}}{\sin^8\theta} 
 					\exp\left(-\tfrac{4}{3}\pi \lambda \frac{\rho^{3}}{\sin^3\theta}\right) ,
\]
\[
  f_{\max}(\phi) =
  \begin{cases}
   \begin{aligned}
      \tfrac{8}{\pi^2} & \sin^2\phi \, [(3\phi - \pi)(3 - 2\sin^2\phi) \\
      		& - (9 - 16\sin^4\phi)\sin\phi\cos\phi]
   \end{aligned}
   &  \phi \in [\frac{\pi}{3}, \frac{\pi}{2} ]  \\
   \begin{aligned}
      \tfrac{8}{\pi^2}\sin^2\phi \, [&(\pi - \phi)(3-2\sin^2\phi) \\
                & + 3\cos\phi\sin\phi] 
    \end{aligned}
    &  \phi \in [\frac{\pi}{2}, \pi ]	
  \end{cases}
\]
The expression for $f_{\max}(\phi)$ is due to Muche~\cite[p.399]{OkabeBook}. 
We begin with the $\theta$ integral:
\[
 F(\rho) = \int_0^{\pi}  \frac{\rho^{8}}{\sin^4\theta} (1+ \cos\theta)^2 \exp\left(-\tfrac{4}{3}\pi \lambda \frac{\rho^{3}}{\sin^3\theta}\right) d\theta . 
\]
An expression for $F(\rho)$ may be given (using Mathematica) in terms of Meijer G-functions and these are then integrated with respect to $\rho$ to find 
\[
\begin{split}
H(\alpha,\phi) = {}& \frac{315}{256}(\tfrac{4}{3}\pi\lambda)^{3} 
		\int_{\alpha}^{\alpha/\sin\phi} F(\rho) d\rho  \\
  = {}& \frac{-35}{128\sqrt{3}} \left[ 6 Z^3 M_1(Z^2) 
  		+ Z^3M_2(Z^2) \right]_{\alpha}^{\alpha/\sin\phi} 
  \end{split}
\]
Where $Z= (\tfrac{4}{3}\pi\lambda \rho^3)/2$, and $M_1, M_2$ are the Meijer G-functions:
\[ M_1(z) = G^{4,1}_{3,5} \left( z \left| \begin{array}{l}
			{-\tfrac{1}{2}}; -\tfrac{1}{3}, \tfrac{1}{3}  \\ 
      {-\tfrac{1}{2}}, -\tfrac{1}{6},\tfrac{1}{6}, \tfrac{1}{2}; -\tfrac{3}{2} 
       \end{array}  \right. \right) ,
\]
\[ M_2(z) = G^{4,1}_{3,5} \left( z \left| \begin{array}{l}
			{-\tfrac{1}{2}}; \tfrac{1}{3}, \tfrac{2}{3}  \\ 
      {-\tfrac{1}{2}}, -\tfrac{1}{6},\tfrac{1}{6}, \tfrac{1}{2}; -\tfrac{3}{2} 
       \end{array}  \right. \right) .
\]
Meijer G-functions are defined by integrals of Gamma functions~\cite{WolframMathematica}.  The form used within Mathematica is  
\[
\begin{split}
&G^{m,n}_{p,q}  \left( z \left| \begin{array}{c}
					a_1, \ldots, a_p \\
					b_1, \ldots, b_q 
					\end{array}  \right. \right)  = \\
& \quad \frac{1}{2\pi i}  \int _{C} 
 \frac{ \Pi_{j=1}^{m} \Gamma(b_j +s)  \Pi_{j=1}^{n} \Gamma(1 - a_j -s) }
 		{ \Pi_{j=n+1}^{p} \Gamma(a_j +s) \Pi_{j=m+1}^{q} \Gamma(1-b_j -s) }
  z^{-s} \; ds . 
\end{split}
\]
where the contour $C$ divides the complex plane into two unbounded regions and separates the poles of $\Gamma(1-a_i-s)$ and the poles of $\Gamma(b_i+s)$. 

Both $M_1(z)$ and $M_2(z)$ diverge as $z \to 0$.  
The products $Z^3 M_i(Z^2) \to 0$ as $Z \to 0$, however, so we determine Taylor expansions about $Z=0$ for these terms.  The zeroth and first order terms vanish and the second derivative has the value 
\[
   \frac{d^2}{dZ^2} \left[ 6 Z^3 M_1(Z^2) + Z^3M_2(Z^2) \right]_{Z = 0} = 13.8564 = A. 
\]
Thus, to second order in $\eta = \tfrac{4}{3}\pi \lambda \alpha^3$: 
\begin{equation*}
  H(\eta, \phi) \sim \frac{-35}{128\sqrt{3}} \frac{A}{2} \left(\frac{\eta}{2}\right)^2 
  		\left( \frac{1}{\sin^6\phi} - 1\right) .
\end{equation*}
We can now compute the integral with respect to $\phi$ of  $H(\eta,\phi)f_{\max}(\phi)$ and obtain the small $\eta$ limit of
\[ 
\begin{split}
 P_{\tri} (\eta) \; \sim & \; \frac{-35}{128\sqrt{3}} \frac{A \eta^2}{8} \int_{\pi/3}^{\pi/2} \left( \frac{1}{\sin^6\phi} - 1\right) f_{\max}(\phi) d\phi \\
 \; = & \; \frac{35}{128\sqrt{3}} \frac{A}{8} \left(\frac{4}{\pi^2} -  \frac{1}{4} \right) \eta^2 .
\end{split}
\]
The expectation per unit volume of the first Betti number for small $\eta$ is therefore 
\begin{equation*}
   E \beta_1  \sim  \frac{48}{35} \pi^2 \lambda P_{\tri} 
   \sim  \frac{\sqrt{3} A}{64} \left( 4 - \frac{\pi^2}{4} \right) \lambda \eta^2 = 0.5747 \lambda \eta^2.
\end{equation*}
This coefficient is exactly that obtained by comparing the expansions for $\chi$ and $\beta_0$, and agrees well with the value obtained in simulations.

\subsection{Empty tetrahedra in 3D}
\label{PDC_3D2}

Finally, we consider the existence of a 2-cycle in an alpha-shape formed by the four faces of a single Poisson-Delaunay tetrahedron.  The conditions for this to occur are that
\begin{enumerate}
 \item all faces belong to the alpha-shape, i.e. $\rho_i < \alpha$, where $\rho_i$ is the circumradius of the face opposite vertex $X_i$; 
 \item the circumcentre is not covered by the union of balls of radius $\alpha$, i.e. the tetrahedron circumradius satisfies $r > \alpha$; 
 \item the circumcenter must be interior to the tetrahedron.
\end{enumerate}
The conditions 1 and 3 both relate to the angle $\theta_i$, between the outward-pointing normal to a face and a vector from the circumcenter to a vertex on that face.  
The circumradius of face-$i$ is $\rho_i = r \sin\theta_i$, so condition 1 becomes 
$r < \alpha/ \sin\theta_i$.   
The condition for the circumcenter to be interior to the tetrahedron requires that 
$\theta_i < \pi/2 $ for $i=0,1,2,3$.  
Thus an integral expression for an empty tetrahedron is:
\begin{equation*}
	P_{tet} = \int_{\theta_0}^{\pi/2}  \int_{\alpha}^{\alpha/\sin\theta} f_{\max} (\theta) f_R(r) dr d\theta ,
\end{equation*}
where $f_R(r)$ is the marginal pdf for the tetrahedron circumradius defined in (\ref{eq:PDC_3Dpdf}) and $f_{\max}(\theta)$ is an unknown pdf for the largest face-normal--vertex angle of a PDC.  The lower limit, $\theta_0$, is the angle for a regular tetrahedron and is $\theta_0 = \arccos\frac{1}{3} = 70.53^{\circ}$. 

Without knowing $f_{\max}(\theta)$, we can still find the leading order term for $P_{tet}$ in the limit of small $\alpha$. 
Firstly, the indefinite integral for the circumradius is
\begin{alignat*}{2} 
    G(r) = & \int  \frac{3}{2}(\tfrac{4}{3}\pi\lambda)^{3} r^{8} \exp(-\tfrac{4}{3}\pi \lambda r^{3}) dr \\
    	   = &  -\left( \frac{(\tfrac{4}{3}\pi\lambda r^3)^2}{2} + \tfrac{4}{3}\pi\lambda r^{3} + 1 \right) \exp(-\tfrac{4}{3}\pi\lambda r^{3}) \\
	   = & - \left( \frac{x^2}{2} + x + 1 \right)\exp(-x).
\end{alignat*}
where we have simplified notation by using the reduced density,
$x = \tfrac{4}{3} \pi\lambda r^3$. 
The Taylor expansion for small $x$ is
\[
   - (\tfrac{1}{2}x^2 + x+ 1) e^{-x} =  - 1 + \frac{x^3}{6} + O(x^4), 
\]
so that to highest order in $\eta = \tfrac{4}{3} \pi\lambda \alpha^3$, 
\[
   G(\eta,\theta) \sim \frac{\eta^3}{6} \left( \frac{1}{\sin^9\theta} - 1 \right). 
\]
In terms of $P_{tet}$ we have 
\begin{equation} \label{eq:Ptet}
  P_{tet} \sim \frac{\eta^3}{6} \int_{\theta_0}^{\pi/2} 
  	f_{\max}(\theta)\left(\frac{1}{\sin^9\theta} - 1\right) d\theta . 
\end{equation}

\begin{figure}
\begin{center}
\includegraphics[width=\columnwidth]{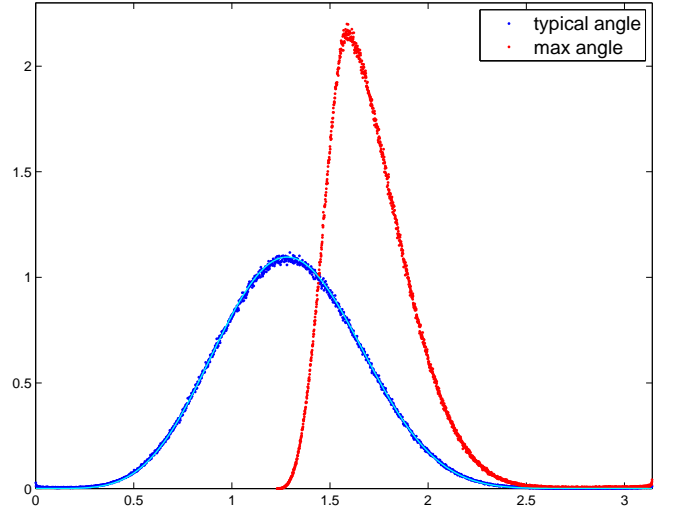}
\caption{Blue dots show the normalised histogram for the typical face-normal -- vertex angle obtained from a simulation of over four million tetrahedra.   The data agree well with the known pdf for this quantity which is plotted as the pale-blue solid curve.  The red dots mark the normalised histogram for the largest face-normal -- vertex angle.  We use this as a numerical approximation to $f_{\max}(\theta)$.}
\label{fig:theta_histograms}
\end{center}
\end{figure}

\begin{figure}
\begin{center}
\includegraphics[width=.85\columnwidth]{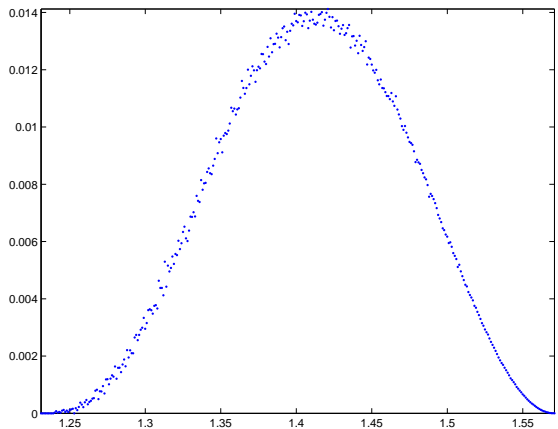}
\caption{Blue dots show the numerical approximation to the integrand in 
Eq.~(\ref{eq:Ptet}) over the domain of integration $\arccos \tfrac{1}{3} < \theta < \tfrac{\pi}{2}$. }
\label{fig:integrand}
\end{center}
\end{figure}

We are unable to derive an analytic expression for $f_{\max}(\theta)$, so we estimate it by simulation and calculate a numerical approximation to $P_{tet}$. 
The Poisson point process is ergodic, so the simplest technique for simulating Poisson-Delaunay cells is to build the Delaunay complex for a large number of points in a cube.  The distribution of tetrahedra in a very large complex is approximately the same as that obtained from many independent realisations.  We generated $10^6$ points with uniform random coordinates in $[-1,1]^3$, built the Delaunay complex, and discarded tetrahedra with circumcenters within a $0.2$ margin of the boundary to minimize edge effects.  This yielded  over four million Poisson-Delaunay tetrahedra.  The probability density for the typical face-normal--vertex angle is known to be \cite{Muche96}
\[
   f (\theta) =  \frac{105}{128}\sin^5\theta(1+\cos\theta)^2 
\]
and provides a check on our simulation.  Normalised histograms for the typical and the largest face-normal--vertex angle in a PDC are shown in Fig.~\ref{fig:theta_histograms}. 
The numerical approximation to the integrand in (\ref{eq:Ptet}) is shown in Fig.~\ref{fig:integrand}.   The area under the curve as calculated from this data is 0.0023.  
Now, since the intensity of Delaunay cells is $\lambda_3 = \frac{24}{35} \pi^2 \lambda$, we have that the expectation per unit volume of the second Betti number in the limit of small $\eta$ is 
\[
E \beta_2(\eta) \sim  \frac{24}{35} \pi^2 \lambda P_{tet} \sim 0.015\, \lambda \eta^3 . 
\] 
Again, we know this result is an asymptotic one because 2-cycles that involve more than the faces of a single Delaunay tetrahedron are necessarily built from five or more overlapping balls.  As we see in Table~\ref{tab:expansions_3d}, the expected number of such clusters has leading order $\eta^4$ as $\eta \to 0$. 

%This coefficient is $20\%$ larger than that estimated from the simulations discussed  Section~\ref{Results}. 

\subsection{Further analysis}

In the study of percolation theory or coverage processes, the total number of connected components is studied via a cluster expansion using expressions for the number of clusters built from $k$-disks~\cite{HallBook,QuintanillaTorquato96}.  A similar approach may be possible for the number of holes in the 2D Poisson-sphere model.  In Section~\ref{PDC_2D} we derived an expression for $P_{\tri}$ the number of holes bounded by three edges.  The next term to consider is $P_{\Box}$, the number of holes bounded by $k=4$ edges.  An expression for $P_{\Box}$ would require a joint distribution for two neighbouring Poisson-Delaunay triangles.  Alternatively, it may be possible to adapt Miles' work \cite{Miles70} on Poisson-generated $k$-figures to study this expansion for $k$-bounded holes in the Poisson alpha-shape.

\begin{acknowledgments}
The author thanks Professor Klaus Mecke for many useful discussions and hospitality over the past few years during the author's visits at the University of Wuppertal, the MPI Metallforschung in Stuttgart, and at the University of Erlangen. 
\end{acknowledgments}

%\appendix
%\section{ }

\bibliography{PoissonBetti}

\begin{thebibliography}{26}
\expandafter\ifx\csname natexlab\endcsname\relax\def\natexlab#1{#1}\fi
\expandafter\ifx\csname bibnamefont\endcsname\relax
  \def\bibnamefont#1{#1}\fi
\expandafter\ifx\csname bibfnamefont\endcsname\relax
  \def\bibfnamefont#1{#1}\fi
\expandafter\ifx\csname citenamefont\endcsname\relax
  \def\citenamefont#1{#1}\fi
\expandafter\ifx\csname url\endcsname\relax
  \def\url#1{\texttt{#1}}\fi
\expandafter\ifx\csname urlprefix\endcsname\relax\def\urlprefix{URL }\fi
\providecommand{\bibinfo}[2]{#2}
\providecommand{\eprint}[2][]{\url{#2}}

\bibitem[{\citenamefont{Mecke and Stoyan}(2002)}]{MeckeBook2}
\bibinfo{editor}{\bibfnamefont{K.}~\bibnamefont{Mecke}} \bibnamefont{and}
  \bibinfo{editor}{\bibfnamefont{D.}~\bibnamefont{Stoyan}}, eds.,
  \emph{\bibinfo{title}{Morphology of Condensed Matter: Physics and Geometry of
  Spatially Complex Systems}}, vol. \bibinfo{volume}{600} of
  \emph{\bibinfo{series}{Lecture Notes in Physics}}
  (\bibinfo{publisher}{Springer}, \bibinfo{year}{2002}).

\bibitem[{\citenamefont{Robins}(2000)}]{RobinsThesis}
\bibinfo{author}{\bibfnamefont{V.}~\bibnamefont{Robins}}, Ph.D. thesis,
  \bibinfo{school}{University of Colorado at Boulder} (\bibinfo{year}{2000}).

\bibitem[{\citenamefont{Robins}(2002)}]{RobinsChapter02}
\bibinfo{author}{\bibfnamefont{V.}~\bibnamefont{Robins}}, in
  \emph{\bibinfo{booktitle}{Morphology of Condensed Matter}}, edited by
  \bibinfo{editor}{\bibfnamefont{K.}~\bibnamefont{Mecke}} \bibnamefont{and}
  \bibinfo{editor}{\bibfnamefont{D.}~\bibnamefont{Stoyan}}
  (\bibinfo{publisher}{Springer}, \bibinfo{year}{2002}), vol.
  \bibinfo{volume}{600} of \emph{\bibinfo{series}{Lecture Notes in Physics}},
  pp. \bibinfo{pages}{261--275}.

\bibitem[{\citenamefont{Blanchard et~al.}(2006)\citenamefont{Blanchard,
  Dobrovolny, Gandolfo, and Ruiz}}]{Blanchard06}
\bibinfo{author}{\bibfnamefont{P.}~\bibnamefont{Blanchard}},
  \bibinfo{author}{\bibfnamefont{C.}~\bibnamefont{Dobrovolny}},
  \bibinfo{author}{\bibfnamefont{D.}~\bibnamefont{Gandolfo}}, \bibnamefont{and}
  \bibinfo{author}{\bibfnamefont{J.}~\bibnamefont{Ruiz}}
  (\bibinfo{year}{2006}), \bibinfo{note}{ar{X}iv:cond-mat/0601344}.

\bibitem[{\citenamefont{Stoyan et~al.}(1995)\citenamefont{Stoyan, Kendall, and
  Mecke}}]{StoyanBook}
\bibinfo{author}{\bibfnamefont{D.}~\bibnamefont{Stoyan}},
  \bibinfo{author}{\bibfnamefont{W.}~\bibnamefont{Kendall}}, \bibnamefont{and}
  \bibinfo{author}{\bibfnamefont{J.}~\bibnamefont{Mecke}},
  \emph{\bibinfo{title}{Stochastic Geometry and its Applications}}
  (\bibinfo{publisher}{Wiley}, \bibinfo{year}{1995}), \bibinfo{edition}{2nd}
  ed.

\bibitem[{\citenamefont{Edelsbrunner et~al.}(1983)\citenamefont{Edelsbrunner,
  Kirkpatrick, and Seidel}}]{Edels83}
\bibinfo{author}{\bibfnamefont{H.}~\bibnamefont{Edelsbrunner}},
  \bibinfo{author}{\bibfnamefont{D.}~\bibnamefont{Kirkpatrick}},
  \bibnamefont{and} \bibinfo{author}{\bibfnamefont{R.}~\bibnamefont{Seidel}},
  \bibinfo{journal}{IEEE Transactions on Information Theory}
  \textbf{\bibinfo{volume}{29}}, \bibinfo{pages}{551} (\bibinfo{year}{1983}).

\bibitem[{\citenamefont{Edelsbrunner and M\"ucke}(1994)}]{Edels94}
\bibinfo{author}{\bibfnamefont{H.}~\bibnamefont{Edelsbrunner}}
  \bibnamefont{and} \bibinfo{author}{\bibfnamefont{E.}~\bibnamefont{M\"ucke}},
  \bibinfo{journal}{ACM Transactions on Graphics}
  \textbf{\bibinfo{volume}{13}}, \bibinfo{pages}{43} (\bibinfo{year}{1994}).

\bibitem[{\citenamefont{Edelsbrunner}(1995)}]{Edels95}
\bibinfo{author}{\bibfnamefont{H.}~\bibnamefont{Edelsbrunner}},
  \bibinfo{journal}{Discrete and Computational Geometry}
  \textbf{\bibinfo{volume}{13}}, \bibinfo{pages}{415} (\bibinfo{year}{1995}).

\bibitem[{\citenamefont{Munkres}(1984)}]{MunkresAT}
\bibinfo{author}{\bibfnamefont{J.}~\bibnamefont{Munkres}},
  \emph{\bibinfo{title}{Elements of Algebraic Topology}}
  (\bibinfo{publisher}{Benjamin Cummings}, \bibinfo{year}{1984}).

\bibitem[{\citenamefont{Delfinado and Edelsbrunner}(1995)}]{DE95}
\bibinfo{author}{\bibfnamefont{C.}~\bibnamefont{Delfinado}} \bibnamefont{and}
  \bibinfo{author}{\bibfnamefont{H.}~\bibnamefont{Edelsbrunner}},
  \bibinfo{journal}{Computer Aided Geometric Design}
  \textbf{\bibinfo{volume}{12}}, \bibinfo{pages}{771} (\bibinfo{year}{1995}).

\bibitem[{\citenamefont{{Duke University BioGeometry group}}()}]{3DAlphaShapes}
\bibinfo{author}{\bibnamefont{{Duke University BioGeometry group}}},
  \emph{\bibinfo{title}{Alpha shapes software}},
  \urlprefix\url{http://biogeometry.cs.duke.edu/software/alphashapes/index.htm%
l}.

\bibitem[{\citenamefont{Clarkson}()}]{HullCode}
\bibinfo{author}{\bibfnamefont{K.}~\bibnamefont{Clarkson}},
  \emph{\bibinfo{title}{A program for convex hulls}},
  \urlprefix\url{http://cm.bell-labs.com/netlib/voronoi/hull.html}.

\bibitem[{\citenamefont{{Tran Kai Frank Da}}(2006)}]{cgal:d-as3-06}
\bibinfo{author}{\bibnamefont{{Tran Kai Frank Da}}}, in
  \emph{\bibinfo{booktitle}{CGAL-3.2 User and Reference Manual}}, edited by
  \bibinfo{editor}{\bibnamefont{{CGAL Editorial Board}}}
  (\bibinfo{year}{2006}).

\bibitem[{\citenamefont{{CGAL Editorial Board}}()}]{CGALurl}
\bibinfo{author}{\bibnamefont{{CGAL Editorial Board}}},
  \emph{\bibinfo{title}{{C}omputational {G}eometry {A}lgorithms {L}ibrary}},
  \urlprefix\url{http://www.cgal.org/}.

\bibitem[{\citenamefont{De~Fabritiis and Coveney}(2003)}]{Coveney03}
\bibinfo{author}{\bibfnamefont{G.}~\bibnamefont{De~Fabritiis}}
  \bibnamefont{and} \bibinfo{author}{\bibfnamefont{P.}~\bibnamefont{Coveney}},
  \bibinfo{journal}{Computer Physics Communications}
  \textbf{\bibinfo{volume}{153}}, \bibinfo{pages}{209} (\bibinfo{year}{2003}).

\bibitem[{\citenamefont{Hall}(1988)}]{HallBook}
\bibinfo{author}{\bibfnamefont{P.~G.} \bibnamefont{Hall}},
  \emph{\bibinfo{title}{Introduction to the theory of coverage processes}}
  (\bibinfo{publisher}{Wiley}, \bibinfo{year}{1988}).

\bibitem[{\citenamefont{Quintanilla et~al.}(2000)\citenamefont{Quintanilla,
  Torquato, and Ziff}}]{QuintTorqZiff00}
\bibinfo{author}{\bibfnamefont{J.}~\bibnamefont{Quintanilla}},
  \bibinfo{author}{\bibfnamefont{S.}~\bibnamefont{Torquato}}, \bibnamefont{and}
  \bibinfo{author}{\bibfnamefont{R.}~\bibnamefont{Ziff}},
  \bibinfo{journal}{Journal of Physics A} \textbf{\bibinfo{volume}{33}},
  \bibinfo{pages}{L399} (\bibinfo{year}{2000}).

\bibitem[{\citenamefont{Quintanilla and
  Torquato}(1996)}]{QuintanillaTorquato96}
\bibinfo{author}{\bibfnamefont{J.}~\bibnamefont{Quintanilla}} \bibnamefont{and}
  \bibinfo{author}{\bibfnamefont{S.}~\bibnamefont{Torquato}},
  \bibinfo{journal}{Physical Review E} \textbf{\bibinfo{volume}{54}},
  \bibinfo{pages}{5331} (\bibinfo{year}{1996}).

\bibitem[{\citenamefont{Lorenz and Ziff}(2001)}]{LorenzZiff01}
\bibinfo{author}{\bibfnamefont{C.}~\bibnamefont{Lorenz}} \bibnamefont{and}
  \bibinfo{author}{\bibfnamefont{R.}~\bibnamefont{Ziff}},
  \bibinfo{journal}{Journal of Chemical Physics}
  \textbf{\bibinfo{volume}{114}}, \bibinfo{pages}{3659} (\bibinfo{year}{2001}).

\bibitem[{\citenamefont{Rintoul}(2000)}]{Rintoul00}
\bibinfo{author}{\bibfnamefont{M.}~\bibnamefont{Rintoul}},
  \bibinfo{journal}{Physical Review E} \textbf{\bibinfo{volume}{62}},
  \bibinfo{pages}{68} (\bibinfo{year}{2000}).

\bibitem[{\citenamefont{Miles}(1974)}]{Miles74}
\bibinfo{author}{\bibfnamefont{R.}~\bibnamefont{Miles}}, in
  \emph{\bibinfo{booktitle}{Stochastic Geometry}}, edited by
  \bibinfo{editor}{\bibfnamefont{E.}~\bibnamefont{Harding}} \bibnamefont{and}
  \bibinfo{editor}{\bibfnamefont{D.}~\bibnamefont{Kendall}}
  (\bibinfo{publisher}{Wiley}, \bibinfo{year}{1974}), chap.
  \bibinfo{chapter}{3.4}, pp. \bibinfo{pages}{202--227}.

\bibitem[{\citenamefont{Okabe et~al.}(2000)\citenamefont{Okabe, Boots,
  Sugihara, and Chiu}}]{OkabeBook}
\bibinfo{author}{\bibfnamefont{A.}~\bibnamefont{Okabe}},
  \bibinfo{author}{\bibfnamefont{B.}~\bibnamefont{Boots}},
  \bibinfo{author}{\bibfnamefont{K.}~\bibnamefont{Sugihara}}, \bibnamefont{and}
  \bibinfo{author}{\bibfnamefont{S.}~\bibnamefont{Chiu}},
  \emph{\bibinfo{title}{Spatial Tessellations: concepts and applications of
  {V}oronoi diagrams}} (\bibinfo{publisher}{Wiley}, \bibinfo{year}{2000}),
  \bibinfo{edition}{2nd} ed.

\bibitem[{\citenamefont{Muche}(1996)}]{Muche96}
\bibinfo{author}{\bibfnamefont{L.}~\bibnamefont{Muche}},
  \bibinfo{journal}{Journal of Statistical Physics}
  \textbf{\bibinfo{volume}{84}}, \bibinfo{pages}{147} (\bibinfo{year}{1996}).

\bibitem[{\citenamefont{Muche}(1998)}]{Muche98}
\bibinfo{author}{\bibfnamefont{L.}~\bibnamefont{Muche}},
  \bibinfo{journal}{Mathematische Nachrichten} \textbf{\bibinfo{volume}{191}},
  \bibinfo{pages}{247} (\bibinfo{year}{1998}).

\bibitem[{\citenamefont{Wolfram}(2003)}]{WolframMathematica}
\bibinfo{author}{\bibfnamefont{S.}~\bibnamefont{Wolfram}},
  \emph{\bibinfo{title}{The Mathematica Book}} (\bibinfo{publisher}{Wolfram
  Media}, \bibinfo{address}{Champaign, IL}, \bibinfo{year}{2003}),
  \bibinfo{edition}{5th} ed.

\bibitem[{\citenamefont{Miles}(1970)}]{Miles70}
\bibinfo{author}{\bibfnamefont{R.}~\bibnamefont{Miles}},
  \bibinfo{journal}{Mathematical Biosciences} \textbf{\bibinfo{volume}{6}},
  \bibinfo{pages}{85} (\bibinfo{year}{1970}).

\end{thebibliography}

\end{document}